\documentclass[prb,aps,twocolumn,amstex,preprintnumbers,amsmath,amssymb,array,tabularx]{revtex4}

\usepackage{graphicx}
\usepackage{dcolumn}
\usepackage{bm}

\DeclareGraphicsExtensions{.jpg,.pdf,.eps}

\begin{document}
\title{A Fully Self-Consistent Treatment of Collective Fluctuations in
Quantum Liquids}
 
\author{Eran Rabani}
\address{ School of Chemistry, The Sackler Faculty of Exact Sciences,
Tel Aviv University, Tel Aviv 69978, Israel}

\author{and \\ {~}\\ David R. Reichman}
\address{ Department of Chemistry and Chemical Biology, Harvard
University, Cambridge, MA 02138 }

\date{\today}

\begin{abstract}
The problem of calculating collective density fluctuations in quantum
liquids is revisited.  A fully quantum mechanical self-consistent
treatment based on a quantum mode-coupling theory [E. Rabani and
D.R. Reichman, {\em J. Chem. Phys.} {\bf 116}, 6271 (2002)] is
presented.  The theory is compared with the maximum entropy analytic
continuation approach and with available experimental results.  The
quantum mode-coupling theory provides semi-quantitative results for
both short and long time dynamics.  The proper description of long
time phenomena is important in future study of problems related to the
physics of glassy quantum systems, and to the study of collective
fluctuations in Bose fluids.
\end{abstract}

\maketitle
 
\newpage 

\section{Introduction}
\label{sec:intro}
The study of the dynamical properties of quantum liquids has a storied
history.\cite{Pinesbook} The interplay of nuclear dynamics, particle
statistics, dimensionality, disorder and temperature can lead to, or
suppress, dramatic effects such as superfluidity in $^{4}\mbox{He}$
and superconductivity in the superfluid Fermi liquid state of
conducting electrons in metals.\cite{Mahanbook} Due to the great
computational difficulties presented by a direct assault on the
real-time many-body Schroedinger equation, microscopic approaches to
calculate the frequency dependent response in condensed phase
disordered quantum systems are approximate and generally rely on
somewhat uncontrolled approximations.

One approach that has been useful in a variety of physical contexts is
the analytic continuation of numerically exact imaginary time
path-integral data.\cite{Gubernatis96,Krilov01c} This approach has
been fruitful for the study of quantum impurity models such as the
Anderson Hamiltonian,\cite{Gubernatis91} as well as models of
correlated electrons such as the Hubbard Hamiltonian.\cite{Pike95}
Berne and coworkers have successfully applied the maximum entropy
(MaxEnt) version of analytic continuation to study the dynamics of
electrons in simple liquids,\cite{Gallicchio94,Gallicchio96}
vibrational and electronic relaxation of impurities in liquid and
solid hosts,\cite{Gallicchio98,Egorov97a} and adiabatic rate constants
in condensed phase environments.\cite{Rabani00,Krilov01b} Boninsegni
and Ceperley have applied MaxEnt to study the dynamic structure factor
of liquid $^{4}\mbox{He}$ above and below the
$\lambda$-transition.\cite{Ceperley96} For the normal state, they find
good agreement with experimental results, while the agreement is
significantly worse in the superfluid state.  In particular, sharp
quasiparticle peaks are not well resolved with the MaxEnt
approach.\cite{Ceperley96} This failure arises from the intrinsic
ill-posed nature of the numerical continuation to the real-time axis.
The conditioning of the data which helps alleviate numerical
instabilities makes the differentiation of fine energy scales
difficult, resulting in smoothing of the response functions.

In several recent
papers,\cite{Rabani02a,Reichman01a,Rabani02b,Reichman02a,Rabani02d} we
have explored a molecular hydrodynamic approach for calculating the
dynamic response functions of quantum liquids.  In particular, we have
formulated a detailed generalization of the classical mode-coupling
approach to the quantum case.\cite{Rabani02b} We have studied both
single particle as well as collective fluctuations, with good
agreement with recent experiments performed on liquid {\em
para}-hydrogen~\cite{Zoppi96a,Bermejo99,Bermejo00,Zoppi01,Zoppi02} and
liquid {\em ortho}-deuterium.\cite{Bermejo93,Zoppi95,Bermejo97} Unlike
the MaxEnt technique, our approach is a {\em theory} and thus provides
additional insight into the dynamics of quantum liquids.  This fact
implies several additional advantages.  First, because our theory is
the direct analog of the approach used in the study of quantum spin
glasses,\cite{Biroli02,Kennett01,Lozano98} interesting connections
between quantum systems with and without quenched disorder may be
made.  For example, our recent study of the spectrum of density
fluctuations in liquid {\em para}-hydrogen, where finite frequency
quasiparticle peaks appear in the dynamic structure factor,
$S(q,\omega)$, resonates with the general result of Cugliandolo and
Lozano that "trivial" quantum fluctuations add coherence to the decay
of dynamics correlators at short-times.\cite{Lozano99} Furthermore, in
principle we could study a range of problems related to the physics of
glassy quantum systems, including the aging behavior of an
out-of-equilibrium quantum liquid.\cite{Biroli02,Kennett01,Lozano98}
Because the MaxEnt approach relies implicitly on the equilibrium
formulation of quantum statistical mechanics, this approach cannot be
used to investigate such questions.

In order to study such interesting problems, as well as the challenges
imposed by nontrivial collective behavior exhibited, for example, by
superfluid $^{4}\mbox{He}$, we need to further develop our theoretical
apparatus.  In our previous studies, we imposed self-consistency in
the study of single-particle properties, but not in the more difficult
case of collective properties such as the dynamics of density
fluctuations.\cite{Rabani02a,Reichman02a,Rabani02d} In this work, we
make several accurate and physically motivated approximations that
render the formal equations presented in Ref. \onlinecite{Rabani02b}
amenable to direct numerical investigation.  We revisit the problem of
the study of $S(q,\omega)$ in liquid {\em ortho}-deuterium and {\em
para}-hydrogen, and show that a fully self-consistent treatment of
density fluctuations greatly improves the agreement between the low
frequency behavior seen in experiments and corrects the behavior of
our previously used quantum viscoelastic
theory.\cite{Rabani02a,Reichman02a,Rabani02d} This low frequency
behavior associated with the long time relaxation of density
fluctuations in these liquids is not well captured by the MaxEnt
approach.

Our paper is organized as follows: In section~\ref{sec:qmct} we
provide an overview of our self-consistent quantum mode-coupling
approach to density fluctuations in quantum liquids.  Furthermore, we
discuss the improvements of the present approach and the physical
approximation that are introduced to make our current study amenable
to path integral Monte Carlo (PIMC) simulation techniques.  In
section~\ref{sec:maxent} we discuss the MaxEnt approach used in the
present study.  Results for collective density fluctuations in liquids
{\em ortho}-deuterium and {\em para}-hydrogen is presented in
section~\ref{sec:results}.  We compare the predictions of our quantum
mode-coupling theory to the MaxEnt results and also to available
experimental results. Finally, in section~\ref{sec:conclusions} we
conclude.

\section{Self-Consistent Quantum Mode-Coupling Theory}
\label{sec:qmct}
In this section we provide a short overview of our quantum
mode-coupling approach suitable for the study of collective density
fluctuations in quantum liquids. For a complete derivation of the
equations described below the reader is referred to
Ref.~\onlinecite{Rabani02b}. The formulation of the quantum
mode-coupling theory (QMCT) described in Ref.~\onlinecite{Rabani02b}
is based on the Kubo transform of the dynamical correlation function
of interest.  For collective density fluctuations the experimental
measured quantity is the dynamic structure factor, $S(q,\omega)$,
which is related to the Kubo transform of the intermediate scattering
function, $F^{\kappa}(q,t)$ by:
\begin{equation}
S(q,\omega) = \frac{\beta \hbar \omega}{2} \left\{
\mbox{coth}\left(\frac{\beta \hbar \omega}{2} \right) + 1 \right\}
S^{\kappa}(q,\omega)
\label{eq:sqw}
\end{equation}
where the Kubo transform of the dynamic structure factor is given in
terms of a Fourier transform of the Kubo transform of the intermediate
scattering function:
\begin{equation}
S^{\kappa}(q,\omega) = \int_{-\infty}^{\infty} dt \mbox{e}^{i \omega
t} F^{\kappa}(q,t).
\label{eq:sqwkubo}
\end{equation}
In the above equations $\beta = \frac{1}{k_{B}T}$ is the
inverse temperature, and the superscript $\kappa$ is a shorthand
notation of the Kubo transform (to be described below).  The quantum
mode-coupling theory described in this section provides a set of
closed, self-consistent equations for $F^{\kappa}(q,t)$, which can be
used to generate the dynamic structure factor using the above
transformations.

We begin with the definition of two dynamical variables, the quantum
collective density operator
\begin{equation}
\hat{\rho}_{\bf q}=\sum_{\alpha=1}^{N} \mbox{e}^{i {\bf q} \cdot {\bf
\hat{r}}_{\alpha}},
\label{eq:rhoq}
\end{equation}
and the longitudinal current operator
\begin{equation}
\hat{j}_{\bf q}=\frac{1}{2m|q|} \sum_{\alpha=1}^{N} \left[({\bf q}
\cdot{\bf \hat{p}}_{\alpha}) \mbox{e}^{i {\bf q} {\bf
\hat{r}}_{\alpha}} + \mbox{e}^{i {\bf q} {\bf \hat{r}}_{\alpha}} ({\bf
\hat{p}}_{\alpha} \cdot{\bf q} ) \right],
\label{eq:jq}
\end{equation}
where ${\bf \hat{r}}_{\alpha}$ is the position vector operator of
particle $\alpha$ with a conjugate momentum ${\bf \hat{p}}_{\alpha}$
and mass $m$, and $N$ is the total number of liquid particles.  The
quantum collective density operator and the longitudinal current
operator satisfy the continuity equation $\dot{\hat{\rho}_{\bf q}} = i
q \hat{j}_{\bf q}$, where the dot denotes a time derivative,
i.e. $\dot{\hat{\rho}_{\bf q}} = \frac{i}{\hbar}
[\hat{H},\hat{\rho}_{\bf q}]$.

Following the projection operator procedure outlined in our early
work,\cite{Rabani02b} the time evolution of the Kubo transform of the
intermediate scattering function, $F^{\kappa}(q,t)=\frac{1}{N} \langle
\hat{\rho}_{\bf q}^{\dagger}, \hat{\rho}_{\bf q}^{\kappa}(t) \rangle$
($\langle \cdots \rangle$ denote a quantum mechanical ensemble
average), is given by the exact quantum generalized Langevin equation
\begin{equation}
\begin{split}
\ddot{F}^{\kappa}(q,t) &+ \omega_{\kappa}^{2}(q) {F}^{\kappa}(q,t)\\ &+
\int_{0}^{t} dt^{\prime}
K^{\kappa}(q,t-t^{\prime})\dot{F}^{\kappa}(q,t^{\prime}) = 0,
\end{split}
\label{eq:fqt}
\end{equation}
where the notation $\kappa$ implies that the quantity under
consideration involves the Kubo transform given by~\cite{Kubo95}
\begin{equation}
\hat{\rho}_{\bf q}^{\kappa} = \frac{1}{\beta \hbar} \int_{0}^{\beta
\hbar} d\lambda \mbox{e}^{-\lambda \hat{H}} \hat{\rho}_{\bf q}
\mbox{e}^{\lambda \hat{H}},
\label{eq:rhokappa}
\end{equation}
and $\hat{H}$ is the Hamiltonian operator of the system.  In the above
equation $\omega_{\kappa}^{2}(q)$ is the Kubo transform of the
frequency factor given by:
\begin{equation}
\omega_{\kappa}^{2n}(q) = \frac{1}{S^{\kappa}(q)} \left <
\frac{d^{n}\hat{\rho}_{\bf q}^{\dagger}}{dt^{n}},
\frac{d^{n}\hat{\rho}_{\bf q}^{\kappa}}{dt^{n}}\right >
\label{eq:omega}
\end{equation}
with $S^{\kappa}(q) = {F}^{\kappa}(q,0) = \frac{1}{N} \langle
\hat{\rho}_{\bf q}^{\dagger},\hat{\rho}_{\bf q}^{\kappa}(0) \rangle$
being the Kubo transform of the static structure factor.  The Kubo
transform of the memory kernel appearing in Eq.~(\ref{eq:fqt}) is
related to the Kubo transform of the random force $\hat{R}_{{\bf
q}}=\frac{d\hat{j}_{\bf q}}{dt} - i |q|
\frac{J^{\kappa}(q)}{S^{\kappa}(q)} \hat{\rho}_{\bf q}$, and is
formally given by
\begin{equation}
K^{\kappa}(q,t) = \frac{1}{N J^{\kappa}(q)} \langle
\hat{R}^{\dagger}_{{\bf q}}, \mbox{e}^{i (1 - {\cal P}^{\kappa}) {\cal
{\hat L}} t}\hat{R}_{{\bf q}}^{\kappa} \rangle,
\label{eq:Fkernel}
\end{equation}
where $J^{\kappa}(q) = \frac{1}{N} \langle \hat{j}_{\bf
q}^{\dagger},\hat{j}_{\bf q}^{\kappa}(0) \rangle$ is the Kubo
transform of the zero time longitudinal current correlation function,
${\cal {\hat L}}=\frac{1}{\hbar} [\hat{H},]$ is the quantum Liouville
operator, and ${\cal P}^{\kappa} = \frac{\langle
\underline{A}^{\dagger},\cdots \rangle} {\langle
\underline{A}^{\dagger}, \underline{A}^{\kappa} \rangle}
\underline{A}^{\kappa}$ is the projection operator used to derive
Eq.~(\ref{eq:fqt}) with the row vector operator given by
$\underline{A} = (\hat{\rho}_{\bf q},\hat{j}_{\bf q})$ [see
Ref.~\onlinecite{Rabani02b} for a complete derivation of
Eqs.~(\ref{eq:fqt}), (\ref{eq:omega}), and (\ref{eq:Fkernel})].

The above equations for the time evolution of the intermediate
scattering function is simply another way for rephrasing the quantum
Wigner-Liouville equation.  The difficulty of numerically solving the
Wigner-Liouville equation for a many-body system is shifted to the
difficulty of evaluating the memory kernel given by
Eq.~(\ref{eq:Fkernel}).  To reduce the complexity of solving
Eq.~(\ref{eq:Fkernel}) we follow the lines of classical mode-coupling
theory~\cite{BalucaniZoppi,BoonYip,HansenMcDonald} to obtain a closed
expression for the memory kernel.  A detailed derivation is provided
elsewhere.\cite{Rabani02b}

We use the approximate form $K^{\kappa}(q,t) = K_{f}^{\kappa}(q,t) +
K_{m}^{\kappa}(q,t)$ where the ``quantum binary'' portion
$K_{f}^{\kappa}(q,t)$ and the quantum mode-coupling portion
$K_{m}^{\kappa}(q,t)$ are obtained using the standard classical
procedure,\cite{BalucaniZoppi,BoonYip,HansenMcDonald} but with a
proper quantum mechanical treatment.  The basic idea behind our
approach is that the binary portion of the memory kernel contains all
the short time information and is obtained from a short time moment
expansion, while the mode-coupling portion of the memory kernel
describes the decay of the memory kernel at intermediate and long
times, and is approximated using a quantum mode-coupling
approach.\cite{Rabani02b}

The fast decaying binary term determined from a short-time expansion
of the exact Kubo transform of the memory function is given by
\begin{equation}
K_{f}^{\kappa}(q,t) = K^{\kappa}(q,0)
\exp\{-(t/\tau^{\kappa}(q))^{2}\}.
\label{eq:KFf}
\end{equation}
The lifetime in Eq.~(\ref{eq:KFf}) is approximated by~\cite{Hirata98}
\begin{equation}
\tau^{\kappa}(q) = [3K^{\kappa}(q,0)/4]^{-1/2}.
\label{eq:tauFf}
\end{equation}
In the above equations $K^{\kappa}(q,0)$ is the zero time moments of
the memory kernel given by
\begin{equation}
K^{\kappa}(q,0) =
\frac{\omega_{\kappa}^{4}(q)}{\omega_{\kappa}^{2}(q)} -
\omega_{\kappa}^{2}(q),
\label{eq:KF0}
\end{equation}
where $\omega_{\kappa}^{2n}(q)$ is given in Eq.~(\ref{eq:omega}).
Note that the exact expression for the lifetime involves higher order
terms such as $\omega_{\kappa}^{6}(q)$ and is given by
$\tau^{\kappa}(q) = [-\ddot{K}^{\kappa}(q,0) /
2K^{\kappa}(q,0)]^{-1/2}$, where $\ddot{K}^{\kappa}(q,0) = -
\omega_{\kappa}^{6}(q) / \omega_{\kappa}^{4}(q) +
(\omega_{\kappa}^{4}(q) / \omega_{\kappa}^{2}(q))^{2}$.  Although the
calculation of $\omega_{\kappa}^{6}(q)$ is possible it involves higher
order derivative of the interaction potential and thus becomes a
tedious task for the path integral Monte Carlo technique.
Interestingly, we find that the approximation to the lifetime given by
Eq.~(\ref{eq:tauFf}) is accurate to within $5$ percent for classical
fluids.\cite{Rabaniunpublished} Thus, we use this simpler approximate
lifetime also for the quantum systems studied here.

The slow decaying mode-coupling portion of the memory kernel,
$K_{m}^{\kappa}(q,t)$, must be obtained from a quantum mode-coupling
approach.  The basic idea behind this approach is that the random
force projected correlation function, which determines the memory
kernel for the intermediate scattering function
(cf. Eq.~(\ref{eq:Fkernel})), decays at intermediate and long times
predominantly into modes which are associated with quasi-conserved
dynamical variables.  It is reasonable to assume that the decay of the
memory kernel at long times will be governed by those modes that have
the longest relaxation time.  The slow decay is basically attributed
to couplings between wavevector-dependent density modes of the form
\begin{equation}
\hat{B}_{{\bf k},{\bf q}-{\bf k}} = \hat{\rho}_{\bf k} {~}
\hat{\rho}_{{\bf q}-{\bf k}},
\label{eq:B}
\end{equation}
where translational invariance of the system implies that the only
combination of densities whose inner product with a dynamical variable
of wavevector $-{\bf q}$ is nonzero is $\hat{\rho}_{\bf k}$ and $
\hat{\rho}_{{\bf q}-{\bf k}}$.

After some tedious algebra the slow mode-coupling portion of the
memory kernel can be approximated by~\cite{Rabani02b}
\begin{equation}
\begin{split}
K_{m}^{\kappa}(q,t) &\approx \frac{2}{(2 \pi)^{3} n J^{\kappa}(q)}
\int d{\bf k} |V^{\kappa}({\bf q},{\bf k})|^{2} \\ \times
&\left[F^{\kappa}(k,t) F^{\kappa}(|{\bf q}-{\bf k}|,t)\right. - \\
&\left.  F_{b}^{\kappa}(k,t) F_{b}^{\kappa}(|{\bf q}-{\bf k}|,t)
\right],
\end{split}
\label{eq:KFm}
\end{equation}
where $n$ is the number density.  In the above equation the binary
term of the Kubo transform of the intermediate scattering function,
$F_{b}^{\kappa}(q,t)$, is obtained from a short time expansion of
$F^{\kappa}(q,t)$ similar to the exact expansion used for the binary
term of $K^{\kappa}(q,t)$, and is given by
\begin{equation}
F_{b}^{\kappa}(q,t) = S^{\kappa}(q) \exp \left\{-\frac{1}{2}
\omega^{2}_{\kappa}(q) t^{2} \right\}.
\label{eq:Fb}
\end{equation}
The subtraction of the product of terms in Eq.~(\ref{eq:KFm})
involving $F_{b}^{\kappa}(q,t)$ is done to prevent over-counting the
total memory kernel at short times, namely, to ensures that the even
time moments of the total memory kernel are exact to forth order in
time.

The vertex in Eq.~(\ref{eq:KFm}) is formally given by
\begin{equation}
\begin{split}
V^{\kappa}({\bf q},{\bf k}) = &\frac{1}{2N} \left( \frac{\langle
\frac{d}{dt}\hat{j}^{\dagger}_{{\bf q}}, \hat{\rho}^{\kappa}_{\bf k}
\hat{\rho}^{\kappa}_{{\bf q}-{\bf k}} \rangle} {S^{\kappa}(k)
S^{\kappa}(|{\bf q}-{\bf k}|)} \right.\\
&\left. -i|q|\frac{J^{\kappa}(q)}{S^{\kappa}(q)} \frac{\langle
\hat{\rho}^{\dagger}_{\bf q} \hat{\rho}_{\bf k}^{\kappa}
\hat{\rho}_{{\bf q}-{\bf k}}^{\kappa}\rangle} {S^{\kappa}(k)
S^{\kappa}(|{\bf q}-{\bf k}|)} \right),
\end{split}
\label{eq:vertex1}
\end{equation}
involving a double Kubo transform.\cite{Reichman00} In the
applications reported below we have calculated the vertex using the
following approximations for the three point correlation functions.
For $\langle \hat{\rho}^{\dagger}_{\bf q} \hat{\rho}_{\bf k}^{\kappa}
\hat{\rho}_{{\bf q}-{\bf k}}^{\kappa}\rangle$ the (Kubo) convolution
approximation has been developed,\cite{Jackson62} leading to:
\begin{equation}
\frac{1}{N}\langle \hat{\rho}^{\dagger}_{\bf q} \hat{\rho}_{\bf
k}^{\kappa} \hat{\rho}_{{\bf q}-{\bf k}}^{\kappa}\rangle \approx S(k)
S^{\kappa}(|{\bf q}-{\bf k}|)S^{\kappa}(q),
\label{eq:convolution}
\end{equation}
while for $\langle \frac{d}{dt}\hat{j}^{\dagger}_{{\bf q}},
\hat{\rho}^{\kappa}_{\bf k} \hat{\rho}^{\kappa}_{{\bf q}-{\bf k}}
\rangle$ we have used the fact that the Kubo transform $J^{\kappa}(q)
= \frac{1}{N} \langle \hat{j}_{\bf q}^{\dagger},\hat{j}_{\bf
q}^{\kappa}(0) \rangle$ can be approximated by $k_{B}T/m$, $m$ being
the mass of the particle, within an error that is less then one
percent for the relevant $q$ values studied in this work.  Based on
this fact, we approximate $\langle \frac{d}{dt}\hat{j}^{\dagger}_{{\bf
q}}, \hat{\rho}^{\kappa}_{\bf k} \hat{\rho}^{\kappa}_{{\bf q}-{\bf k}}
\rangle$ by:
\begin{equation}
\begin{split}
\frac{1}{N}\left< \frac{d}{dt}\hat{j}^{\dagger}_{{\bf q}},
\hat{\rho}^{\kappa}_{\bf k} \hat{\rho}^{\kappa}_{{\bf q}-{\bf k}}
\right > &\approx -\frac{k_{B}T}{m q} \left ({\bf q}\cdot {\bf k}
S^{\kappa}(|{\bf q}-{\bf k}|) \right. \\ & \left. + ({\bf q} -{\bf k})
\cdot {\bf q} S^{\kappa}(k) \right).
\end{split}
\label{eq:3pj}
\end{equation}
The combination of these approximations lead to a simplified vertex
given by
\begin{equation}
\begin{split}
V^{\kappa}({\bf q},{\bf k}) \approx \frac{i k_{B} T}{2 m q} \left(
\frac{{\bf q}\cdot {\bf k}}{S^{\kappa}(k)} + \frac{({\bf q} -{\bf k})
\cdot {\bf q}}{S^{\kappa}(|{\bf q}-{\bf k}|)} - q^{2}
\frac{S(k)}{S^{\kappa}(k)} \right),
\end{split}
\label{eq:vertex}
\end{equation}
where $S(k)$ is the quantum mechanical static structure factor (non
Kubo transformed version).  At high values of $k$ the vertex should
decay to zero, while the approximation fails to do so.  Hence, in the
applications discuss below we employ a cutoff to overcome the
shortcoming of our approximation.  Below we discuss the choice of
cutoff for the two model systems studied.

To obtain the Kubo transform of the intermediate scattering function,
one requires as input the frequency factor $\omega_{\kappa}^{2}(q)$,
and the memory kernel $K^{\kappa}(q,t)$.  Since the memory kernel
depends on $F^{\kappa}(q,t)$, the equation of motion for the
intermediate scattering function (Eq.~(\ref{eq:fqt})) must be solved
self-consistently.  The time-independent terms in the memory kernel,
and the frequency factor can be obtained from {\em static} equilibrium
input using a suitable path integral Monte Carlo
scheme.\cite{Rabani02b,Rabani01a}

\section{Analytic Continuation of the Intermediate Scattering Function}
\label{sec:maxent}
An alternative approach to the quantum mode-coupling theory is based
on the numerical maximum entropy analytic continuation approach, which
has recently been used by Boninsegni and Ceperley to study density
fluctuations in liquid helium.\cite{Ceperley96} In this section we
provide a short outline of maximum entropy analytic continuation
approach applicable to study collective density fluctuations in
quantum liquids.

The analytic continuation of the intermediate scattering function is
based on the Fourier relation between $S(q,\omega)$ and $F(q,t)$:
\begin{equation}
F(q,t) = \frac{1}{2 \pi} \int_{-\infty}^{\infty} d\omega \mbox{e}^{-i
\omega t} S(q,\omega).
\label{eq:fqt1}
\end{equation}
The dynamic structure factor is thus analogous to the spectral density
used in the analytic continuation of spectral line
shapes~\cite{Gallicchio96,Gallicchio94} and to the frequency dependent
rate constant or diffusion constant used in analytic continuation of
rates.\cite{Rabani00,Rabani02c} By performing the replacement $t
\rightarrow -i\tau$, and using the detailed balance relation
$S(q,-\omega) = e^{-\beta \omega} S(q,\omega)$ we obtain
\begin{equation}
\tilde{F}(q,\tau) = \frac{1}{2 \pi} \int_{0}^{\infty} d\omega
\biggl[e^{- \omega \tau} + e^{(\tau-\beta) \omega}
\biggr] S(q,\omega),
\label{eq:fqtau1}
\end{equation}
where $t, \tau \ge 0$, and
\begin{equation}
\tilde{F}(q,\tau) = \frac{1}{Z} \frac{1}{N} \mbox{Tr} \left(e^{-\beta
H} e^{\tau H} \hat{\rho}_{\bf q}^{\dagger} e^{-\tau H} \hat{\rho}_{\bf
q} \right).
\label{eq:fqtau}
\end{equation}
The reason for introducing the imaginary time intermediate scattering
function, $\tilde{F}(q,\tau)$, is that, unlike its real time
counterpart, it is straightforward to obtain it using an appropriate
path-integral Monte Carlo simulation
technique.\cite{Berne86ra,Berne86rb} However, in order to obtain the
dynamic structure factor and the real time intermediate scattering
function one has to invert the integral in Eq.~(\ref{eq:fqtau1}).  Due
to the singular nature of the integration kernel the inversion of
Eq.~(\ref{eq:fqtau1}) is an ill-posed problem.  As a consequence, a
direct approach to the inversion would lead to an uncontrollable
amplification of the statistical noise in the data for
$\tilde{F}(q,\tau)$, resulting in an infinite number of solutions that
satisfy Eq.~(\ref{eq:fqtau1}).  Clearly, in this case, little can be
said about the real time dynamics and the corresponding dynamic
structure factor.

In recent years, Bayesian ideas have been used to deal with the
ill-posed nature of continuing the noisy imaginary time Monte Carlo
data to real time.\cite{Gubernatis96,Krilov01c} One of the most widely
used approaches is the maximum entropy
method.\cite{Krilov01c,Skilling89} The method requires only that the
transformation which relates the data and the solution be known.
Furthermore, maximum entropy allows the inclusion of prior knowledge
about the solution in a logically consistent fashion.  As such, the
method is well-suited for solving ill-posed mathematical problems.

In the language of the maximum entropy method $\tilde{F}(q,\tau)$ is
the data, $K(\tau,\omega)=e^{- \omega \tau} + e^{(\tau-\beta) \omega}$
is the singular kernel, and $S(q,\omega)$ is the solution, also
referred to as the map.  Maximum entropy principles provide a way to
choose the most probable solution which is consistent with the data
through the methods of Bayesian inference.  Typically, the data is
known only at a discrete set of points $\{\tau_{j}\}$, and thus we
search for the solution at a discrete set of points $\{\omega_{l}\}$.
The maximum entropy method selects the solution which maximizes the
posterior probability, or the probability of the solution
$S(q,\omega_{l})$ given a data set $\tilde{F}(q,\tau_{j})$.  The
posterior probability is given by~\cite{Skilling89}
\begin{equation}
{\cal{P}}(S(q,\omega)|\tilde{F}(q,\tau)) \propto \exp(\alpha_{q}
S_{q} - \chi_{q}^{2}/2).
\label{eq:posteriorprob}
\end{equation}
Here $\chi_{q}^2$ is the standard mean squared deviation from the
data, and $S_{q}$ is the information entropy.\cite{Rabani02c}

Obtaining the maximum entropy solution then involves finding a map
$S(q,\omega)$ which maximizes the posterior probability and is
therefore a maximization problem in $M$ variables, where $M$ is the
number of points $\{\omega_{l}\}$ at which the solution is evaluated.
The solution obtained in this way is still conditional on the
arbitrary parameter $\alpha_{q}$, which can be interpreted as a
regularization parameter.  In this work, we use a flat default map
that satisfies a known sum rule, such as the integral over
$S(q,\omega)$, and $\alpha_{q}$ is selected according to the L-curve
method.\cite{Lawson95} In this case we regard $\alpha_{q}$ as a
regularization parameter controlling the degree of smoothness of the
solution, and entropy as the regularizing function.  The value of
$\alpha_{q}$ is selected by constructing a plot of
$\log[-S_{q}(S(q,\omega))]$ vs.  $\log \chi_{q}^{2}$. This curve has a
characteristic L-shape, and the corner of the L, or the point of
maximum curvature, corresponds to the value of $\alpha_{q}$ which is
the best compromise between fitting the data and obtaining a smooth
solution.

\section{Results}
\label{sec:results}
Although it is known that liquid {\em ortho}-deuterium and liquid {\em
para}-hydrogen may be treated as Boltzmann particles without the
complexity of numerically treating particle
statistics,\cite{Sindzingre91} they still exhibits some of the
hallmarks of a highly quantum liquid.  In fact, recent
theoretical~\cite{Rabani02a,Reichman01a,Rabani02b,Reichman02a,Rabani02d,Bermejo00,Zoppi02,Rabani02c,Voth96,Kinugawa98,Zoppi02a}
and experimental
studies~\cite{Zoppi96a,Bermejo99,Bermejo00,Zoppi01,Bermejo97} show
that these dense liquids are characterized by quantum dynamical
susceptibilities which are not reproducible using classical theories.
Thus, these liquids are ideal to assess the accuracy of methods
developed for quantum liquids such as the self consistent quantum
mode-coupling theory and the maximum entropy analytic continuation
approach.

Here we report on a direct comparison between these methods for
collective density fluctuations in liquid {\em ortho}-deuterium and
liquid {\em para}-hydrogen.  A comparison between the two approaches
has been made for self-transport in liquid {\em
para}-hydrogen~\cite{Rabani02c} where good agreement has been observed
for the real time velocity autocorrelation function.  However, the
present comparison is more challenging since the experimental dynamic
structure factor for these liquids is characterized by more than a
single frequency peak (unlike the case of
self-transport),\cite{Bermejo99,Bermejo97} indicating that more than a
single timescale is involved in the relaxation of the density
fluctuations.  So far, in the context of our molecular hydrodynamic
approach, density fluctuations in these dense liquids has been
described only within the quantum viscoelastic
model~\cite{Rabani02a,Reichman02a,Rabani02d} which fails to reproduce
the low frequency peak in $S(q,\omega)$ that is associated with long
time dynamics.  Therefore, one challenge for the improved
self-consistent quantum hydrodynamic approach and for the analytic
continuation method is to recover this long time behavior.

\subsection{Technical details}
To obtain the static input required by the quantum mode-coupling
theory and the imaginary time intermediate scattering function
required for the analytic continuation approach we have performed PIMC
simulations at $T=20.7K$ and $\rho=0.0254\mbox{\AA}^{-3}$ for liquid
{\em ortho}-deuterium~\cite{Zoppi95} and $T=14K$ and
$\rho=0.0235\mbox{\AA}^{-3}$ for liquid {\em
para}-hydrogen.\cite{Scharf93} The PIMC simulations were done using
the NVT ensemble with $256$ particles interacting via the
Silvera-Goldman potential,\cite{Silvera78,Silvera80} where the entire
molecule is described as a spherical particle, so the potential
depends only on the radial distance between particles.  The staging
algorithm~\cite{Pollock84} for Monte Carlo chain moves was employed to
compute the numerically exact input.  The imaginary time interval was
discretized into $P$ Trotter slices of size $\epsilon = \beta / P$
with $P=20$ and $P=50$ for liquid {\em ortho}-deuterium and liquid
{\em para}-hydrogen, respectively.  Approximately $2\times 10^{6}$
Monte Carlo passes were made, each pass consisted of attempting moves
in all atoms and all the beads that were staged.  The acceptance ratio
was set to be approximately $0.25-0.3$ for both liquids.

The static input obtained from the PIMC simulations was then used to
generate the memory kernel ($K^{\kappa}(q,t)=K_{f}^{\kappa}(q,t) +
K_{m}^{\kappa}(q,t)$)) and frequency factor ($\omega_{\kappa}^{2}(q)$)
needed for the solution of the equations of motion for
$F^{\kappa}(q,t)$.  To obtain $K_{m}^{\kappa}(q,t)$ we solved
Eq.~(\ref{eq:KFm}) with a cutoff in $|q-k|$ to overcome the divergent
behavior discussed above (cf. Eq.~(\ref{eq:vertex})).  The choice of
this cutoff is straightforward given that the approximate vertex
decays to zero at intermediated values of $k$ before the unphysical
divergent behavior steps in.  We have used $|q-k|_{cut}=5.66 \AA^{-1}$
and $|q-k|_{cut}=4.73 \AA^{-1}$ for {\em ortho}-deuterium and {\em
para}-hydrogen, respectively.  Since the memory kernel depends on the
value of the $F^{\kappa}(q,t)$ itself the solution must be obtained
self-consistently.  The initial guess for the memory kernel was taken
to be equal to the fast binary portion.  The integro-differential
equations were solved using a fifth-order Gear predictor-corrector
algorithm.\cite{AllenTildesley} Typically, less than $10$ iterations
were required to converge the correlation function, with an average
error smaller than $10^{-8}$ percent.

The same PIMC simulation runs were used to generate the imaginary time
$\tilde{F}^{\kappa}(q,\tau)$ and the corresponding covariance
matrices.  We then used the L-curve method to determine the optimal
value of the regularization parameter, $\alpha_{q}$.  In all the
results shown below the value of $\alpha_{q}$ ranged between $5$ to
$20$ depending on the value of $q$ (the results were not found to be
very sensitive to the value of $\alpha_{q}$ within a reasonable range
$\alpha_{q}=1-50$).  The plots of $\log[-S_{q}(S(q,\omega))]$ vs.
$\log \chi_{q}^{2}$ for all values of $q$ result in a very sharp
L-shape curves, indicating the high quality of the PIMC data.

\subsection{Liquid {\em ortho}-deuterium}
\begin{figure}
\begin{center}
\includegraphics[width=8cm]{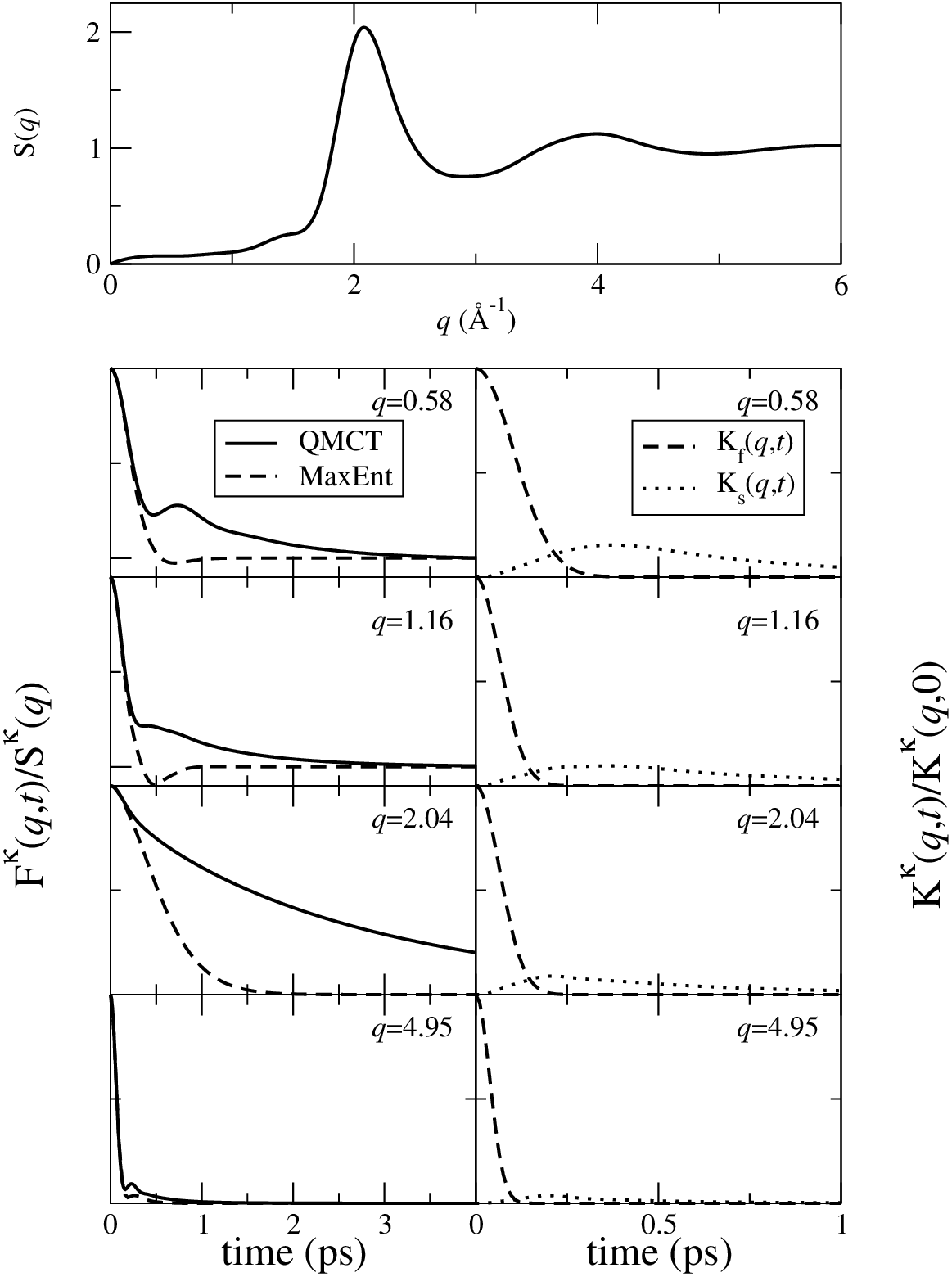}
\end{center}
\caption{Plots of the static structure factor (upper panel), the Kubo
transform of the intermediate scattering function (left panels), and
the Kubo transform of the quantum binary and mode-coupling portions of
the memory kernel (right panels) for liquid {\em ortho}-deuterium.
The values of $q$ indicated in the panels are in units of $\AA^{-1}$.}
\label{fig:fqtod2}
\end{figure}

The results for the Kubo transform of the intermediate scattering
function for liquid {\em ortho}-deuterium obtained from the QMCT and
the MaxEnt method are shown in the left panels of
Fig.~\ref{fig:fqtod2} for several values of $q$.  The right panels of
Fig.~\ref{fig:fqtod2} show the binary and mode-coupling portions of
the memory kernel obtained from the QMCT for the same values of $q$.
The MaxEnt results were generated from $S(q,\omega)$ by taking the
Fourier transform of $S(q,\omega)$ divided by the proper frequency
factor to ``Kubo'' the real time correlation function.

The most striking result is the discrepancy between the QMCT and the
MaxEnt result observed at intermediate and long times for all values
of $q \le q_{max}$, where $q_{max} \approx 2 \AA^{-1}$ is the value of
$q$ where $S(q)$ reaches its first maximum (upper panel of
Fig.~\ref{fig:fqtod2}).  The agreement between the QMCT which is exact
to order $t^{6}$ and the MaxEnt results at short times is expected.
However, the MaxEnt result fails to reproduce the proper oscillatory
behavior observed in $F^{\kappa}(q,t)$ that gives rise to a peak in
$S(q,\omega)$ at a finite frequency (see Fig.~\ref{fig:sqwod2} below),
signifying the existence of collective coherent excitations in liquid
{\em ortho}-deuterium.  Common to both approaches is that as $q$
approaches $q_{max}$ the decay rate of the Kubo transform of the
intermediate scattering function decreases giving rise to a quantum
mechanical de Gennes narrowing of the dynamic structure factor (see
Fig.~\ref{fig:sqwod2} below).  In addition, we observed an increase in
the decay rate of $F^{\kappa}(q,t)$ at high values of $q$, where the
two approaches yield nearly identical results on a sub-picosecond
timescale.  The agreement between the two approaches at high values of
$q$ is not surprising, since the decay of the intermediate scattering
function is rapid, on timescales accessible to the analytic
continuation approach.

The results shown in Fig.~\ref{fig:fqtod2} are the first application
of the self-consistent quantum mode-coupling theory described in
section~\ref{sec:qmct} and suggested in Ref.~\onlinecite{Rabani02b}.
As clearly can be seen in the right panels of Fig.~\ref{fig:fqtod2}
the contribution of the quantum mode-coupling portion of the memory
kernel is significant for all values of $q$ below $q_{max}$, while at
the highest value of $q$ shown the contribution of
$K_{m}^{\kappa}(q,t)$ is negligible.  This is consistent with results
obtained for classical dense
fluids,\cite{BalucaniZoppi,BoonYip,HansenMcDonald} signifying the need
to include the long time portion of the memory kernel.

\begin{figure}
\begin{center}
\includegraphics[width=8cm]{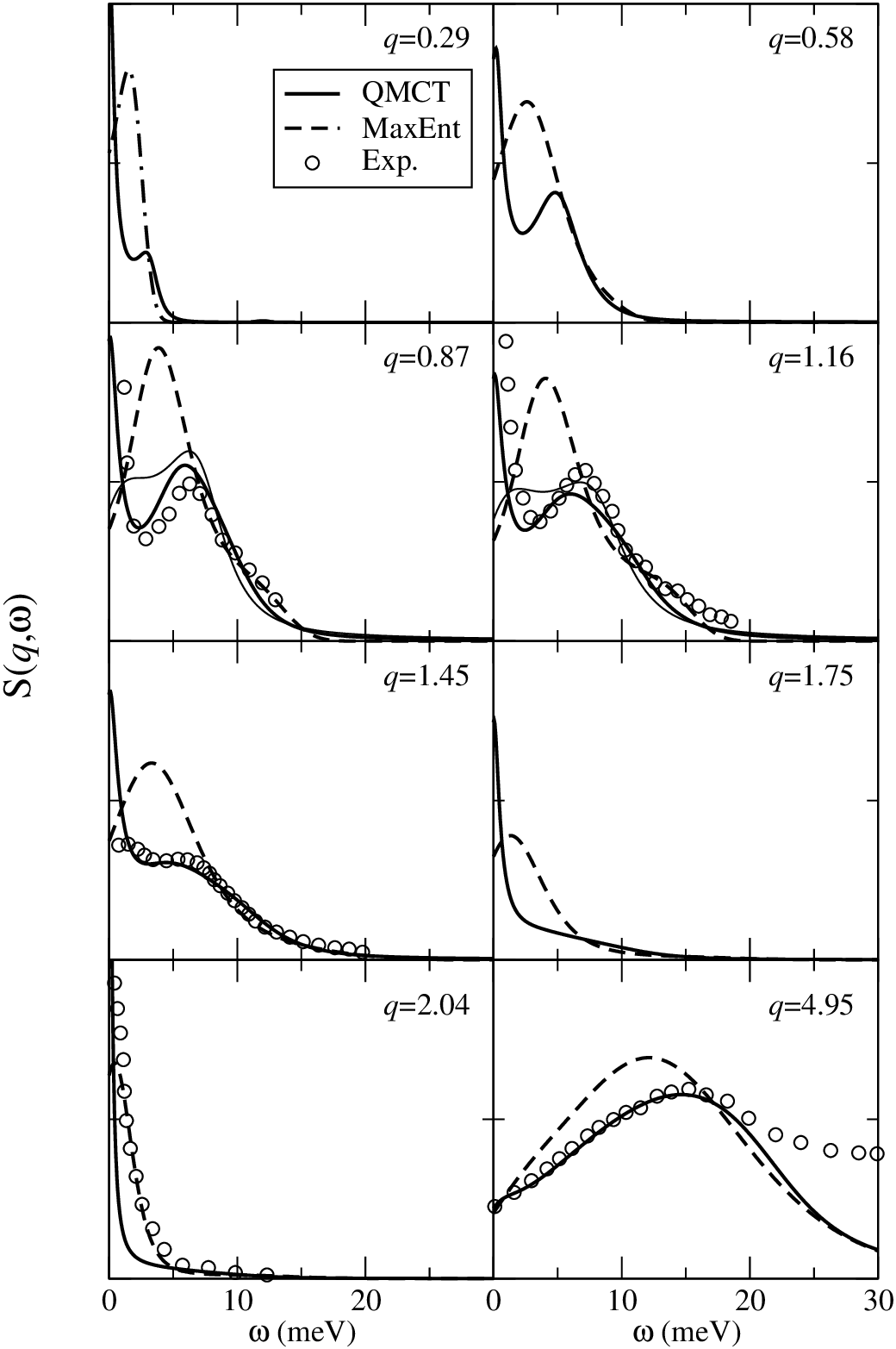}
\end{center}
\caption{A plot of the dynamic structure factor of liquid {\em
ortho}-deuterium for several values of $q$ (in units of $\AA^{-1}$).
The thick solid and dashed lines are QMCT and MaxEnt results ,
respectively.  The thin solid line is the result of
QVM,\cite{Rabani02d} and open circles are the experimental
results.\cite{Bermejo97}.}
\label{fig:sqwod2}
\end{figure}

To quantify the discrepancy between the QMCT and the MaxEnt result for
$F^{\kappa}(q,t)$ we have calculated $S(q,\omega)$ and compared the
results to the experiments of Mukherjee {\em et al.}~\cite{Bermejo97}
for several values of $q$ (the theoretical values of $q$ are slightly
different from the experiments due to the limitations associated with
the constant volume simulation approach).  The results shown in
Fig.~\ref{fig:sqwod2} are normalized such that
$\int_{-\infty}^{\infty} d\omega S(q,\omega) = 1$.  The agreement
between the experimental results~\cite{Bermejo93,Bermejo97} and the
QMCT is excellent.  In particular the theory captures the position of
both the low and high intensity peaks and their width for nearly all
wavevectors shown.  The fact that for limiting cases the QMCT somewhat
underestimates the width of the low intensity peak maybe attributed to
the broadening associated with the instrumental
resolution.\cite{Bermejo97} This is certainly the case at $q_{max}$
where the result of the QMCT, which is in excellent agreement with the
result obtained from QVM (not shown),\cite{Rabani02d} underestimates
the width of the single low frequency peak.  The overall good
agreement between the QMCT and the experiments is remarkable since our
previous study based on the quantum viscoelastic approach failed to
reproduce the low intensity peak associated with long time dynamics
(see thin solid line in Fig.~\ref{fig:fqtod2}).  Thus, the inclusion
of the quantum mode-coupling portion to the memory kernel is important
to properly describe long time phenomena.  In order to test the
accuracy of the approximation for the vertex given by
Eq.~(\ref{eq:vertex}) we have replaced it with its classical
limit.\cite{BalucaniZoppi} The results for the low frequency part of
$S(q,\omega)$ (not shown) are in poor agreement with the experiments.
In particular the low intensity peak is absent at intermediate values
of $q$, and the higher frequency peak is somewhat shifted.

Turning to the MaxEnt results for $S(q,\omega)$, it becomes obvious
why MaxEnt fails to provide a quantitative description of the density
fluctuations in liquid {\em ortho}-deuterium.  It is well known that
the MaxEnt approach fails when several timescales arise in a problem.
This is clearly the case here where the MaxEnt approach predicts a
single frequency peak instead of two, at a position that is
approximately the averaged position of the two experimental peaks.
Only when the dynamics are characterized by a single relaxation time,
like the case at $q_{\max}$, the MaxEnt approach provides quantitative
results.  On the other hand, MaxEnt provides quantitative results for
the short time dynamics as reflected in the width of the finite
frequency peak in $S(q,\omega)$ (see also the discussion in
Fig.~\ref{fig:fqtod2}).

\subsection{Liquid {\em para}-hydrogen}
In many ways liquid {\em para}-hydrogen is very similar to liquid {\em
ortho}-deuterium.  While the interaction potential is the same for
both liquids within the Born-Oppenheimer approximation (for zero
molecular angular momentum), the phase diagram is somewhat different
due to the lower mass of {\em para}-hydrogen.  Therefore, we expect
that the accuracy of the QMCT will be similar to that obtained for
{\em ortho}-deuterium.  As will become clear below this is not the
case.  While qualitative features are captured by QMCT (and not by the
MaxEnt analytic continuation approach), the agreement between the
experiments and the theory is far from the quantitative agreement
observed for liquid {\em ortho}-deuterium.  

\begin{figure}
\begin{center}
\includegraphics[width=8cm]{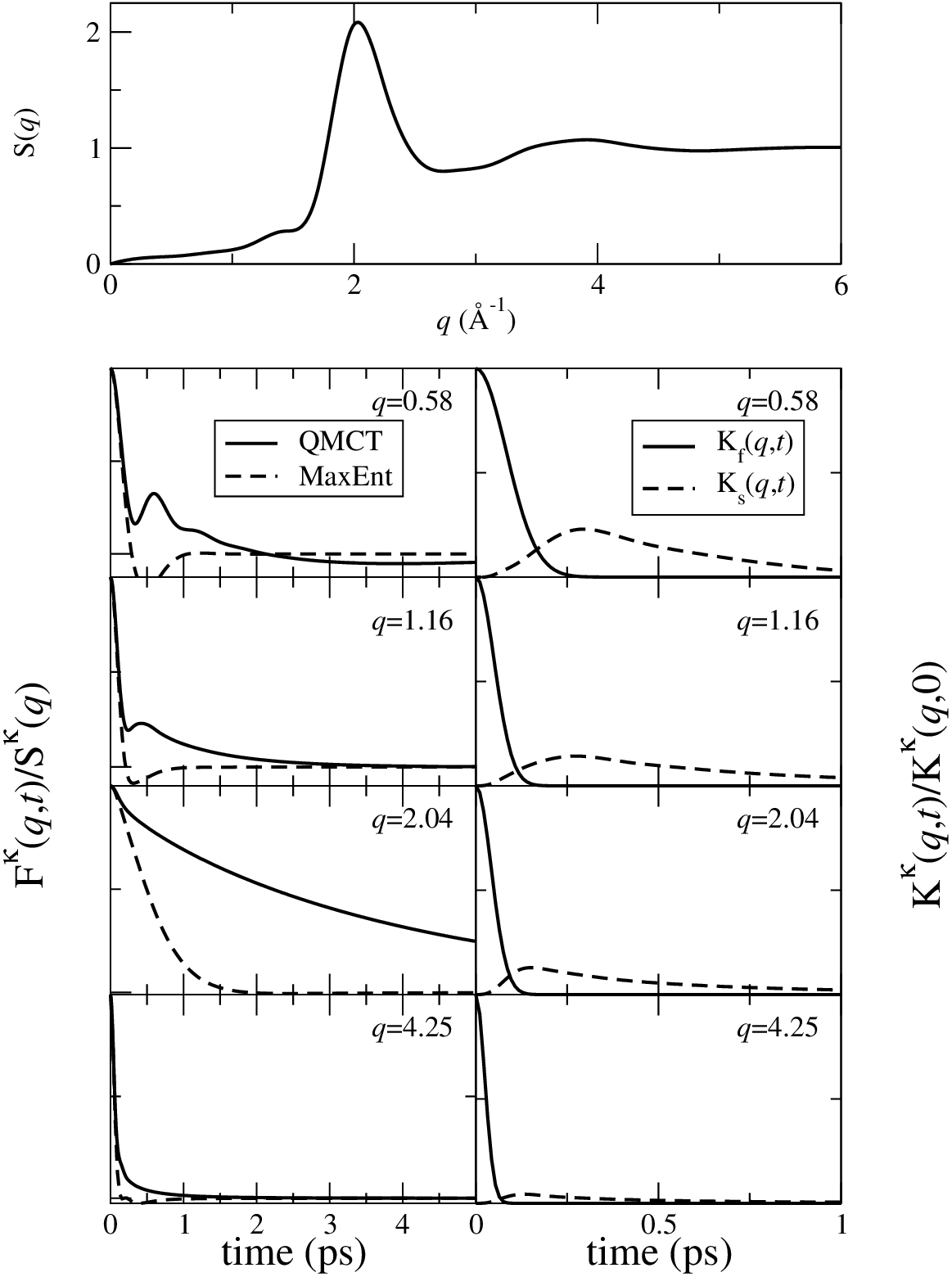}
\end{center}
\caption{Plots of the static structure factor (upper panel), the Kubo
transform of the intermediate scattering function (left panels), and
the Kubo transform of the quantum binary and mode-coupling portions of
the memory kernel (right panels) for liquid {\em para}-hydrogen.  The
values of $q$ indicated in the panels are in units of $\AA^{-1}$.}
\label{fig:fqtph2}
\end{figure}

In Fig.~\ref{fig:fqtph2} we show the results for the Kubo transform of
the intermediate scattering function for liquid {\em para}-hydrogen
obtained from the QMCT and the MaxEnt method (left panels) for several
values of $q$.  The right panels of Fig.~\ref{fig:fqtph2} show the
binary and mode-coupling portions of the memory kernel obtained from
the QMCT for the same values of $q$.  The agreement between the QMCT
and the MaxEnt result is good at short times.  Both approaches capture
the quantum mechanical de Gennes narrowing associated with the
decrease in the decay rate of the Kubo transform of the intermediate
scattering function as $q$ approaches $q_{max}$.  In addition both
capture the increase in the decay rate of $F^{\kappa}(q,t)$ at high
values of $q$ where the two yield nearly identical results.  However,
similar to the case of liquid {\em ortho}-deuterium, at intermediate
and long times when $q \le q_{max}$ the results deviate markedly from
each other.

\begin{figure}
\begin{center}
\includegraphics[width=8cm]{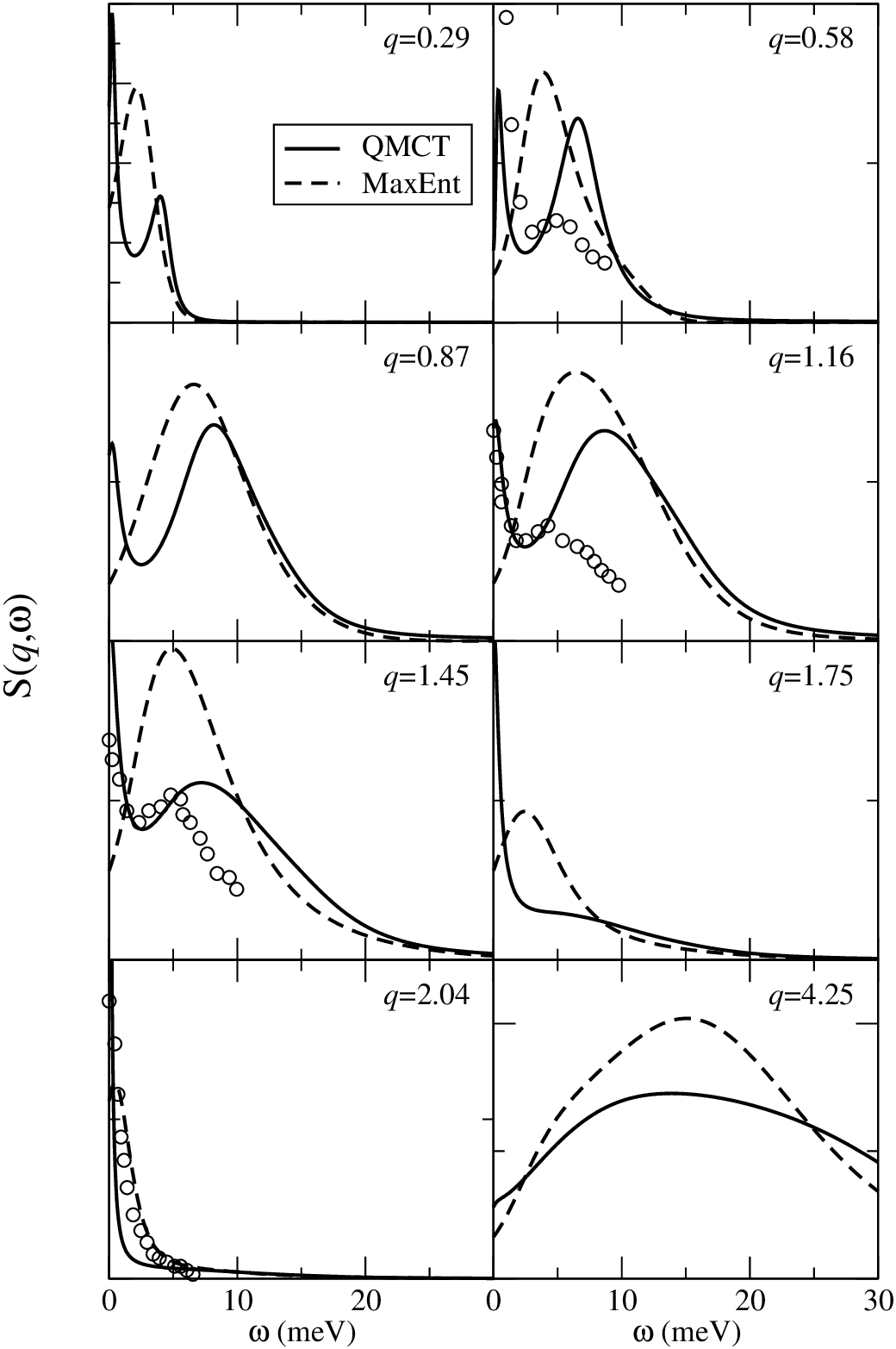}
\end{center}
\caption{A plot of the dynamic structure factor of liquid {\em
para}-hydrogen for several values of $q$ (in units of $\AA^{-1}$).
The solid and dashed lines are QMCT and MaxEnt results ,
respectively. Open circles are the experimental results of Bermejo
{\em et al.}\cite{Bermejo99}}
\label{fig:sqwph2}
\end{figure}

The QMCT results show pronounced oscillatory behavior in
$F^{\kappa}(q,t)$ that gives rise to two peaks in $S(q,\omega)$ at a
finite frequency as can be seen in Fig.~\ref{fig:sqwph2}.  Only one
peak is observed in the MaxEnt approach.  Similar peaks were also
observed experimentally by Bermejo {\em et al.}\cite{Bermejo99} and
using the centroid molecular dynamics (CMD)~\cite{Voth96review}
approach by Kinugawa.\cite{Kinugawa98} As can be seen in the figure,
the experimental result for the higher frequency peak is slightly
shifted to lower frequencies with a smaller amplitude.  This
discrepancy between the QMCT and the experiments is not completely
clear. One missing factor is the instrument response function, which
we have not included.  We would expect that this will lower the
amplitude of the peaks that we have calculated.  It is also possible
that, along with inaccuracy incurred by our approach, there may be
some difficulties in extracting the experimental
response.\cite{Zoppi02a} It is interesting to note that the
discrepancy of peak location is similar to that produced by the CMD
result.  However, the use of CMD for this problem is unjustified
because the density operator is not a linear function of phase space
coordinates.

Finally, we would like to note that the agreement between the QMCT and
the MaxEnt method is excellent for the self-transport in {\em
para}-hydrogen,\cite{Rabani02c} while the results for the dynamic
structure factor plotted in Fig.~\ref{fig:sqwph2} show significant
deviations between the two approaches.  Most likely the discrepancy
results from multiple frequency peaks in $S(q,\omega)$.  MaxEnt is
nearly quantitative as long as $S(q,\omega)$ is characterized by a
single peak.  Namely, at high $q$ values where the decay of
$F^{\kappa}(q,t)$ is on a sub-picosecond timescale, and at $q$ values
near $q_{max}$.

\section{Conclusions}
\label{sec:conclusions}
The problem of calculating collective density fluctuations in quantum
liquids has been revisited.  Two techniques have been applied to study
the dynamic structure factor in liquids {\em ortho}-deuterium and {\em
para}-hydrogen: The self-consistent quantum mode-coupling theory and
the numerical maximum entropy analytic continuation approach. We find
that the results obtained using the QMCT for collective density
fluctuations are in excellent agreement with the experiments on liquid
{\em ortho}-deuterium for a wide range of $q$ values.  On the other
hand the results obtained using the MaxEnt approach deviate from the
experiments for nearly the entire relevant wavevector range.  Failure
of the MaxEnt result was attributed to the presence of {\em two} peaks
in $S(q,\omega)$.  Improvements of the MaxEnt, such as including real
time dynamics in the inversion of the singular integral, may provide
more accurate results.

The excellent agreement between our QMCT and the experiments for
liquid {\em ortho}-deuterium is an encouraging indicator of the
accuracy of the QMCT.  The fact that our QMCT captures the position of
both the low and high intensity peaks in the dynamic structure factor
and their width for nearly all wavevectors studied is an important
result, since our previous studies in which we invoked a single
relaxation time for the memory kernel of the QGLE (the QVM) failed to
reproduce the low frequency
peak.\cite{Rabani02a,Reichman02a,Rabani02d} It is the self-consistent
treatment of the mode-coupling portion of the memory kernel that
accounts for a proper description of the intermediate to long time
dynamics is these systems, and the proper treatment of the vertex in
the memory kernel.

The agreement between the QMCT and the experimental results for liquid
{\em ortho}-deuterium emphasize the need for new experiments on liquid
{\em para}-hydrogen.  But liquid {\em para}-hydrogen is not the only
system that can be addressed using the QMCT.  Future work in other
directions, including the dynamic properties of liquid helium, is
currently underway.

\section{Acknowledgments}
\label{sec:acknowledgments}
This work was supported by The Israel Science Foundation founded by
the Israel Academy of Sciences and Humanities (grant number 31/02-1 to
E.R.). D.R.R. is an Alfred P. Sloan Foundation Fellow and a Camille
Dreyfus Teacher-Scholar.



\pagebreak
\end{document}